\documentclass[]{aastex}

\usepackage{emulateapj5}
\usepackage{onecolfloat}
\usepackage{graphicx} 
\usepackage{fancyheadings} 
\usepackage{ulem}
\usepackage{rotating}
\usepackage{lscape}

\newcommand{\ew}{W_\lambda}
\newcommand{\feh}{[{\rm Fe/H}]}

\newcommand{\etal}{et al.\ }

\newcommand{\lya}{Ly$\alpha$ }

\newcommand{\cm}[1]{\, {\rm cm^{#1}}}
\newcommand{\N}[1]{{N({\rm #1})}}

\newcommand{\rAA}{{\AA \enskip}}

\newcommand{\smm}{\sum\limits}

\begin{document}

\twocolumn[%
\submitted{Accepted to the Astrophysical Journal December 22, 2000}

\title{GALACTIC CHEMICAL ABUNDANCES AT $z>3$ I: First Results from the 
Echellette Spectrograph and Imager}

\author{ JASON X. PROCHASKA\altaffilmark{1}}
\affil{The Observatories of the Carnegie Institute of Washington}
\affil{813 Santa Barbara St. \\
Pasadena, CA 91101}
\email{xavier@ociw.edu}
\and
\author{ERIC GAWISER\altaffilmark{1} \& ARTHUR M. WOLFE\altaffilmark{1}}
\affil{Department of Physics, and Center for Astrophysics and Space Sciences}
\affil{University of California, San Diego; 
C--0424; La Jolla, CA 92093}
\email{awolfe@ucsd.edu,egawiser@ucsd.edu}

\begin{abstract} 

We present the first results from an ongoing survey to discover and measure
the metallicity of $z>3$ damped \lya systems with the Echellette 
Spectrograph and Imager (ESI) on the Keck~II telescope.  
Our motivation arises from a recent study on the damped \lya systems
suggesting only mild evolution in the cosmic metallicity from 
$z \sim 2$ to 4.  The Echellette Spectrograph and Imager, which provides
two complementary spectroscopic modes, is the ideal instrument for a
$z>3$ damped \lya survey.  We describe our observing strategy and report
on the discovery and analysis of 5 new $z>3$ damped \lya systems acquired
in a single night of observing.   These observations further support the
principal conclusions of the previous study: (1) the cosmic metallicity in
neutral gas inferred from the damped \lya systems does not evolve 
significantly from $z \sim 2 $ to 4; (2) the unweighted metallicity exhibits
a statistically significant decrease with increasing redshift; 
and (3) not a single damped
\lya system has a metallicity below $\lbrack$Fe/H$\rbrack$~$= -3$.  
We discuss the implications of these
results and comment on recent theoretical studies which attempt to 
explain the observations.


\end{abstract}

]
\altaffiltext{1}{Visiting Astronomer, W.M. Keck Telescope.
The Keck Observatory is a joint facility of the University
of California and the California Institute of Technology.}

\pagestyle{fancyplain}
\lhead[\fancyplain{}{\thepage}]{\fancyplain{}{PROCHASKA, GAWISER \& WOLFE}}
\rhead[\fancyplain{}{GALACTIC CHEMICAL ABUNDANCES AT $z>3$}]{\fancyplain{}{\thepage}}
\setlength{\headrulewidth=0pt}
\cfoot{}

\section{INTRODUCTION}

Determining the chemical enrichment history places fundamental constraints
on the processes of galaxy formation.  Both the stellar and gas  
metallicity are diagnostics of a number of physical processes including:
the star formation rate, metal transport, the 
initial mass function, and metal production yields.
For several decades, researchers have investigated the chemical enrichment
history of the universe by examining in great detail the chemical evolution 
of our Galaxy \citep{chiosi80,chiapp97,boissier99}.  Specifically,
one examines the age-metallicity relation of various Galactic stellar
populations and extrapolates these results to the universe as a whole under
the assumption that our Galaxy is not overly peculiar.
This extrapolation not withstanding, the approach is limited by the 
large uncertainties in measuring stellar ages and metallicities, particularly
in the most metal-poor and presumably oldest stars \citep{edv93}.
While it is possible to obtain spectroscopy of individual stars for several
nearby galaxies \citep[e.g.][]{smecker99,shetrone00},
metallicities of external galaxies are 
usually determined by examining spectra 
of the integrated light \citep{trager00}.  
Unfortunately this approach is severely limited
by the well-known age, metallicity, dust degeneracy and pursuing metallicity
measurements beyond the local universe is extremely challenging.
At high redshift ($z\sim3$), \cite{ptt00b} have introduced a method for
estimating the metallicity of the Lyman break galaxies, but unfortunately
they have applied it to only a single galaxy as the analysis requires a 
relatively bright galaxy in order to obtain a moderate-resolution, high
signal-to-noise spectrum.

Quasar absorption line studies, which examine the gas properties of galaxies,
allow an independent and complementary method
for studying chemical evolution.  In particular,
studies of the damped \lya systems -- neutral hydrogen gas layers with
HI column densities in excess of $2 \times 10^{20} \cm{-2}$ --
provide a measure of the mean metallicity of the universe {\it in neutral gas}
from $z$ = 0 to 5 \citep{ptt94,ptt97,pei95,pro00}. 
Because the damped \lya systems are detected in absorption they are
believed to present a more representative sample of protogalaxies at high
redshift (i.e.\ probing the entire galactic mass function)
than galaxies observed in emission \citep[e.g.]{kau96,grd97,mmw98}.
With observations of the damped \lya systems, one
directly studies the mean galactic enrichment as a function of time and
can test scenarios of chemical enrichment \citep[e.g.][]{pei99}.
Furthermore, measurements of the relative abundances of elements like Si, Zn,
Mn, Fe \citep{lu96,pro96,pro97,pro99}
provide an investigation into the star formation rate and initial mass function
of these protogalaxies.

\begin{table*}[ht]
\begin{center}
\caption{
{\sc NEW DLA DISCOVERED WITH ESI} \label{tab:obs}}
\begin{tabular}{lccccccc}
\tableline
\tableline
QSO & $z_{em}$& $R$ & $z_{abs}$ & $\N{HI}$ & $N_{mtl}$ & [Fe/H] & Ion \\
\tableline
PSS0808+52 & 4.45 & 18.82 & 3.114 & $20.59 \pm 0.10$ 
	& $14.15 \pm 0.05$ & $-1.94 \pm 0.11$ & Fe~II 2374\\
PSS0957+33 & 4.25 & 17.59 & 3.279 & $20.32 \pm 0.08$   
	& $14.18 \pm 0.03$ & $-1.62 \pm 0.09$ & Fe~II 2344\\
& & & 4.178 & $20.50 \pm 0.10$   
	& $13.84 \pm 0.07$  & $-2.16 \pm 0.13$ & Fe~II 1608\\
PSS1248+31 & 4.35 & 18.9 & 3.696 & $20.43 \pm 0.10$   
	& $13.89 \pm 0.05$  & $-2.04 \pm 0.11$ & Fe~II 1608\\
PSS1432+39 & 4.28 & 18.6 & 3.272 & $20.95 \pm 0.10$   
	& $15.10 \pm 0.10$ & $-1.35 \pm 0.15$ & Ni~II 1751, Cr~II 2056\\
\tableline
\end{tabular}
\end{center}
\end{table*}

In a recent paper (Prochaska \& Wolfe 2000; hereafter PW00) 
we presented evidence that the mean 
metallicity of the damped \lya systems exhibits no significant evolution
from $z = 2$ to 4.  This result is in contradiction with the majority of 
chemical evolution models all of which significantly
underpredict the observed enrichment at $z>3$ (e.g.\ Edmunds \& Phillips 1997;
Pei et al.\ 1999; Mathlin et al.\ 2000; 
but see Cen \& Ostriker 1999; Prantzos \& Boissier 2000 as discussed below).
For this epoch, however, the observational result was
based on [Fe/H] measurements of only $\sim 10$ 
protogalaxies and therefore suffered from small number statistics.
Observationally, it is very expensive to analyze $z>3$ damped \lya systems
with high resolution spectrographs (e.g.\ HIRES, UVES) because the majority
of known $z>3$ damped \lya systems lie towards faint quasars.
To address the uncertainty arising from the small sample size at $z>3$, 
we have initiated a program using the new Echellette Spectrograph and
Imager on the Keck~II telescope.
The project takes advantage of the unique capabilities of this instrument.
The combination of high efficiency, moderately high 
resolution ($R \approx 10000$), and complete wavelength coverage from 
$\lambda \sim 4000 - 11000$~\rAA provide an ideal match to a $z>3$ survey
of damped \lya systems.  
In this Letter we present our first results from the survey, specifically
metallicity measurements for five damped \lya 
systems discovered during one night at Keck Observatory.  
We compare these measurements against the results from PW00 and discuss
the implications for chemical enrichment in the early universe.  In a future
paper, we will present the complete set of
observations and investigate
chemical abundance patterns of elements like Si, Fe, Al, and Ni.

\section{OBSERVATIONS AND REDUCTION}
\label{sec-obs}

The Echellette Spectrograph and Imager (ESI; PI: Joe Miller, UCSC) 
is mounted at the cassegrain focus of the W.M. Keck~II 10m telescope.  
The instrument
has two spectrographic modes: (1) a prism-dispersed low resolution mode
with resolution $R$ decreasing linearly from $R=4800$ at 3900~\rAA to 
$R=1000$ at $1.1\mu$; and (2) a multi-order echellette mode with 
$R \approx 10000$ over the entire spectrum ($\approx 11$~km/s/pix).  
Both modes have continuous wavelength coverage from
$\lambda = 4000 - 11000$~\AA. 

We observed the 4 quasars listed in Table~\ref{tab:obs} on the night
of April 7, 2000.  All of them were drawn from the
Palomar Sky Survey \citep{djg98} kindly made public by G. Djorgovski and
collaborators\footnote{http://www.astro.caltech.edu/$\sim$george/dposs/index.html}.
At the time of the observations, none
of the quasars were known to exhibit damped \lya systems.
We acquired 400s exposures of each quasar in the low dispersion mode with a
$1''$ slit and reduced the data in real
time with the IRAF task {\it apall} and related tasks. We then
obtained two or three 1200s exposures in the echellette 
mode (0.75$''$ slit lending resolution FWHM $\approx 45$~km/s) 
of those quasars with
promising damped \lya candidates; i.e., a \lya line with rest
equivalent width $W_\lambda > 10$~\rAA \citep[e.g.][]{wol95}.  
In this fashion we were able to 
identify a significant number of new $z>3$ damped \lya systems and measure
their metallicity in the same night of observing.  
Because the quasars were previously unobserved at a resolution sufficient
for a damped \lya study, this
survey will also help improve the statistics of the cosmological neutral
gas density at $z>3$ \citep{storr00,peroux00}.  

We reduced the echellette observations with the echelle suite of
IRAF tasks, in part following a recipe developed by S. 
Castro\footnote{http://www.astro.caltech.edu/$\sim$smc/esi.html}.
After subtracting the overscan and dividing by a normalized quartz
flat, we traced the curved orders of a bright standard star, determined
the offset for each faint object, and extracted the data with {\it apall}.
We then extracted a CuArHgXe arc-lamp spectrum for each quasar and 
determined a wavelength solution using {\it ecidentify}.  Finally,
the data were flux calibrated by correcting the observed flux to our
observations of the spectrophotometric standard Hiltner 600 
\citep{massey90}, and normalized
to unit flux using a package similar to the IRAF task {\it continuum}.
In the future, we intend to reduce the data with a modified version of
the {\it makee} packaged developed by T. Barlow for HIRES echelle 
data\footnote{http://spider.ipac.caltech.edu/staff/tab/makee/}.
For most of the observations, the S/N per pixel is $\approx 20$ redward
of \lya emission decreasing to $\sim 10$ at $\lambda > 9000$~\rAA and
$\lambda < 4500$~\AA.

\begin{figure}[ht]
\includegraphics[height=3.7in, width=3.2in,angle=-90]{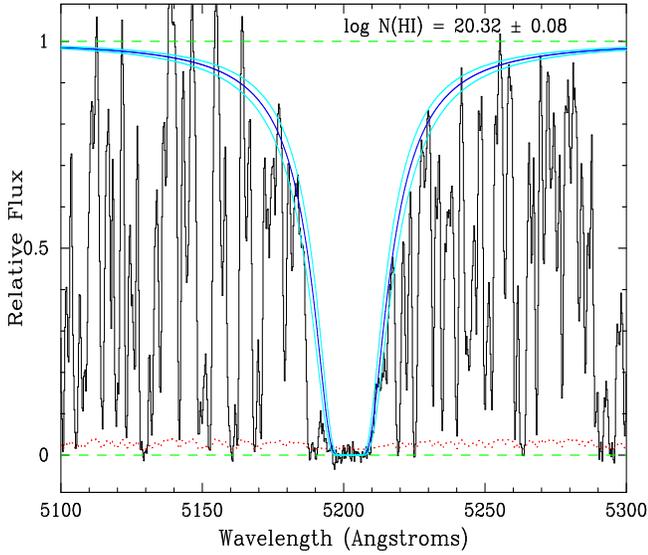}
\caption{\lya profile for the damped \lya system at $z=3.279$ towards
PSS0957+33.  The dark solid line traces the best fit value of
$\N{HI} = 20.32$ and the light solid lines show
an estimated 1$\sigma$ error of 0.08~dex.}
\label{fig:lya}
\end{figure}

\section{METALLICITY ANALYSIS AND RESULTS}
\label{sec-analy}

We searched each of the normalized echellette spectra for potential 
damped \lya systems and fitted the \lya profiles with Voigt profiles
to estimate their HI column densities $\N{HI}$.  At a resolution 
$R \sim 10000$,
the typical \lya forest cloud is just resolved and we could account
for the majority of contaminating features in the wings of the \lya profile.
As an example, 
Figure~\ref{fig:lya} presents the \lya profile for the damped \lya system
at $z= 3.279$ towards PSS0957+33.  
The dark curve overplotted on the data is a Voigt profile with
$\log \N{HI} = 20.32$ and the light grey
lines represent a conservative $1\sigma$
error estimate of $\pm 0.08$~dex.  The $\N{HI}$ values and $1\sigma$ errors
for each of the damped \lya systems are listed in column 5
of Table~\ref{tab:obs}.

With a FWHM resolution of $\approx 45$~km/s, we had concerns that the
absorption lines from some systems would be unresolved and our analysis
would underestimate the metallicity.  To quantify this potential systematic
error, we analyzed an ESI echellette spectrum of the very bright quasar
PH957.  There is a known damped \lya system at $z=2.3$ towards PH957
with previous 
HIRES observations \citep{wol94,pro99} which reveal that the
metal-line profiles exhibit the
majority of absorption within a 20~km/s interval, i.e.\ well within the
ESI resolution.  We compared the column density
measurements determined from the apparent optical depth method
\citep{sav91} from 10 absorption lines observed with both ESI and HIRES.
For absorption lines with $\ew< 0.5$~\rAA all of the column density values
agreed to within the $2 \sigma$ statistical error and $80\%$ were within 
$1 \sigma$.  Even several obviously saturated profiles yielded column
densities to within 0.2~dex.  Therefore, 
we expect that transitions with $\ew< 0.5$~\rAA will yield 
column density measurements free of saturation and we have focused on
these transitions in our analysis.

\begin{figure}[hb]
\includegraphics[height=5.4in, width=3.8in]{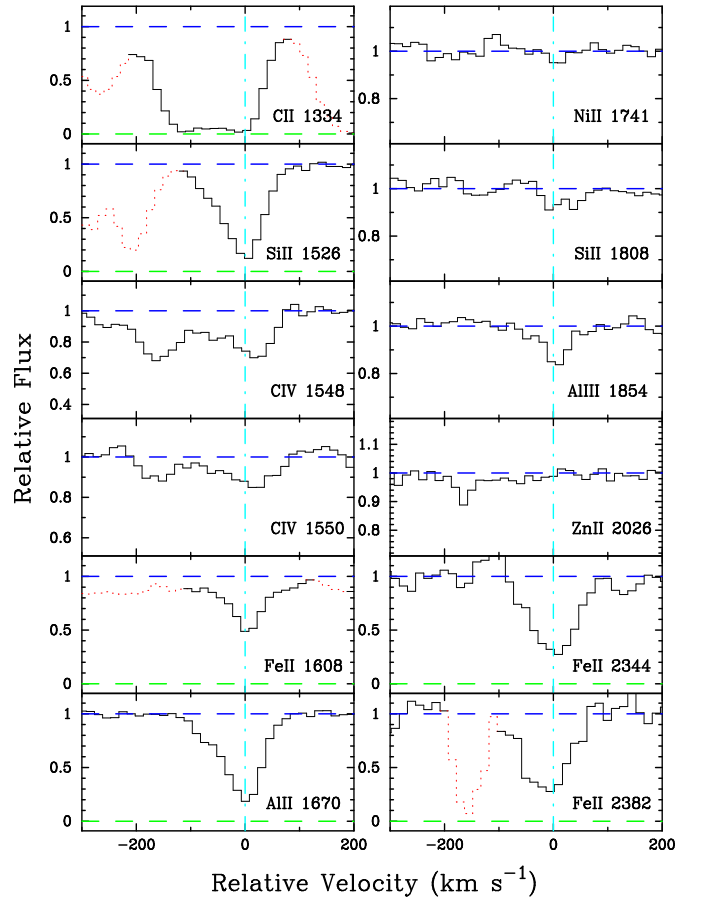}
\caption{Metal-line transitions observed for the damped \lya system at
$z=3.279$ towards PSS0957+33.  The Fe$^+$ column density is well constrained
by measurements of the Fe~II $\lambda 1608, 2344, 2832$ profiles using
the apparent optical depth method.
}
\label{fig:mtl}
\end{figure}

Depending on the absorption redshift, the HI column density, and the
metallicity of the damped system, the absorber will exhibit various metal-line
profiles which can be analyzed to assess its metallicity
(e.g.\ Figure~\ref{fig:mtl}).
Ideally, we measured an Fe$^+$ column density for each system from the 
Fe~II $\lambda$ 1608,1611,2344,2382 transitions.  
These measurements were checked for
consistency against the $\N{Si^+}$, $\N{Ni^+}$, $\N{Al^+}$ observations
by considering the relative abundance patterns typically observed for
the damped \lya systems \citep{lu96,pro99}.  
In one case (PSS1432+39) all of the
observed Fe~II profiles were blended with 
significant absorption features or too weak/strong to provide a reliable
column density measurement. To estimate
the metallicity we relied on the observations of other ions,
in particular Ni$^+$ and Cr$^+$ which show very nearly solar relative
abundances (i.e.\ [Ni/Fe]=0) in nearly every damped \lya system.  For this
system, we adopted an additional 0.05~dex error.  The metallicities and
errors are presented in Table~\ref{tab:obs} for each system.  We also
indicate the most important metal-line transition(s) used in determining the 
metallicity.

Because we have relaxed our definition of the Fe metallicity to include
systems with observed Ni, Al, or Cr transitions, we now include several
other systems from our echelle database 
(BR0019$-$15, BRI1346$-$03; Prochaska \& Wolfe 1999).  Following our practice
for the ESI observations, we adopt an additional 0.05~dex error for these
metallicities.  We have also revised our [Fe/H] estimates for 
the systems towards Q1215+33 and Q0000$-$26 as the former system had 
an [Fe/H] value based on a saturated Fe~II 1608 profile and we believe
the Fe abundance for Q0000$-$26 is more reliably given by [Ni/H] which agrees
well with the Fe abundance derived by \cite{molaro00}.
Finally, we also include two new observations from the literature:
the UVES observations of J0307$-$4945 
\citep{miro00} and the measurement the damped \lya system towards
Q0201+11 \citep{ellison00}.  
Table~\ref{tab:echsum} summarizes all of the echelle measurements.
Even with our relaxed criteria, we exclude
a single damped \lya system at $z=4.203$ towards BRI0951$-$04 for which
we have only detected Si~II transitions, [Si/H]~$= -2.56$ \citep{pro99}.  
Because this system exhibits
a low HI column density, its inclusion would not 
significantly affect any of the following
analysis except for the observed scatter in [Fe/H] at $z>3$.

\begin{table}[ht]
\begin{center}
\caption{
{\sc DLA ECHELLE METALLICITIES \label{tab:echsum}}}
\begin{tabular}{lccccc}
\tableline
\tableline
QSO & $z_{abs}$ & $\log \N{HI}$ & [Fe/H] & Ref\tablenotemark{e} \\
\tableline
Q1331+17&1.776&21.18&$-2.077 \pm  0.041$ & 1\\  
Q2230+02&1.864&20.85&$-1.162 \pm  0.086$ & 1\\  
Q2206-19&1.920&20.65&$-0.732 \pm  0.073$ & 2\\  
Q1215+33\tablenotemark{a} &1.999&20.95&$-1.604 \pm  0.122$ & 1\\  
Q0458-02&1.999&21.65&$-1.642 \pm  0.102$ & 1\\  
Q2231-002&2.066&20.56&$-1.309 \pm  0.100$ & 1\\  
Q2206-19&2.076&20.43&$-2.634 \pm  0.062$ & 2\\  
Q2359-02&2.095&20.70&$-1.693 \pm  0.103$ & 1\\  
Q0149+33&2.141&20.50&$-1.799 \pm  0.102$ & 1\\  
Q0528-2505\tablenotemark{b} &2.141&20.70&$-1.270 \pm  0.140$ & 3\\  
Q2359-02&2.154&20.30&$-1.905 \pm  0.105$ & 1\\  
Q0216+08&2.239&20.45&$-1.060 \pm  0.183$ & 3\\  
Q2348-14&2.279&20.56&$-2.266 \pm  0.077$ & 1\\  
PH957&2.309&21.40&$-1.804 \pm  0.065$ & 1\\  
Q0841+12\tablenotemark{a} &2.375&20.95&$-1.648 \pm  0.142$ & 1\\  
Q2343+12&2.431&20.34&$-1.228 \pm  0.100$ & 4\\  
Q0201+36&2.463&20.38&$-0.900 \pm  0.045$ & 5\\  
Q1223+17\tablenotemark{d} &2.466&21.50&$-1.723 \pm  0.104$ & 6\\  
Q0841+12&2.476&20.78&$-1.846 \pm  0.101$ & 1\\  
Q2344+12&2.538&20.36&$-1.951 \pm  0.103$ & 4\\  
Q1759+75&2.625&20.80&$-1.225 \pm  0.108$ & 1\\  
Q1425+6039&2.827&20.30&$-1.320 \pm  0.058$ & 3\\  
Q0347-38&3.025&20.80&$-1.828 \pm  0.100$ & 1\\  
Q1055+46&3.317&20.34&$-1.902 \pm  0.101$ & 4\\  
Q0201+11&3.386&21.26&$-1.410 \pm  0.113$ & 7\\  
Q0000-2619\tablenotemark{a} &3.390&21.41&$-2.226 \pm  0.142$ & 8\\  
BR0019-15\tablenotemark{a} &3.439&20.92&$-1.461 \pm  0.159$ & 1\\  
BRI1108-07&3.608&20.50&$-2.144 \pm  0.101$ & 6\\  
BRI1346-03\tablenotemark{c} &3.736&20.72&$-2.663 \pm  0.153$ & 1\\  
BRI0951-04&3.857&20.60&$-2.038 \pm  0.117$ & 1\\  
BR2237-0607&4.080&20.52&$-2.170 \pm  0.167$ & 3\\  
BRI0952-01&4.203&20.55&$-1.996 \pm  0.126$ & 6\\  
PSS1443+27&4.226&20.80&$-0.971 \pm  0.114$ & 6\\  
BR1202-07&4.383&20.60&$-2.220 \pm  0.188$ & 3\\  
J0307-4945&4.466&20.67&$-1.940 \pm  0.333$ & 9\\  
\tableline
\end{tabular}
\end{center}
\tablenotetext{a}{[Fe/H] estimated from Ni}
\tablenotetext{b}{[Fe/H] estimated from Cr}
\tablenotetext{c}{[Fe/H] estimated from Al}
\tablenotetext{d}{Note the
tabulated metallicity for the system towards Q1223+17 is erroneous in PW00.
The correct value, however, was used in all of the figures and
analysis of PW00.} 
\tablenotetext{e}{Key to References -- 1: \cite{pro99}; 2: \cite{pro97};
3: \cite{lu96}; 4: \cite{lu97}; 5: \cite{pro96}; 6: \cite{pro00};
7: \cite{ellison00}; 8: \cite{molaro00}; 9: \cite{miro00}}
\end{table}

\section{DISCUSSION}
\label{sec-discuss}

Figure~3 presents the [Fe/H] metallicity measurements
for $z>1.7$ against absorption redshift\footnote{This redshift corresponds
to the optical atmospheric cutoff for the detection of the \lya transition}.
The open squares are measurements
derived from echelle observations.
and the new ESI measurements are presented as stars. 
For all data points,
the size of the data points is linear with the log of the HI column density.
Following PW00, we plot the column density weighted mean metallicity
$<Z> \; \equiv \, \log(\Omega_{metals}/\Omega_{gas}) - \log Z_\odot =
\log [\smm \N{Fe^+} / \smm \N{HI}] - 
\log ( {\rm Fe/H} )_\odot$ for two epochs: 
$z_{low} = [1.5,3]$ and $z_{high} = (3,4.5]$ with median
$\hat z_{low} = 2.20$,
$\hat z_{high} = 3.65$.  
We have calculated a 
statistical error in $<Z>$ based on standard error propagation and 
estimated the uncertainty due to sample variance with a bootstrap
analysis (PW00).  We find $<Z>_{low} = -1.53 \pm 0.04$ (statistical error)
with a bootstrap 
error of 0.09~dex and $<Z>_{high} = -1.63 \pm 0.04$ with bootstrap
error 0.12~dex.  Similarly the unweighted logarithmic means 
$<\feh>$ for the two
samples with bootstrap errors only are $<\feh>_{low} = -1.58 \pm 0.10$ and 
$<\feh>_{high} = -1.89 \pm 0.07$.
Finally, the scatter in the two samples is 
$\sigma(\feh)_{low} = 0.46$ and $\sigma(\feh)_{high} = 0.40$ each with
a bootstrap error of 0.07.
Comparing the three statistical moments for the two epochs, we note
the values for the $z_{high}$ sample
are in excellent agreement with those from PW00 (there is no change for
the $z_{low}$ sample).  As discussed in PW00, there is no significant 
evolution in $<Z>$ yet a decrease in $<\feh>$ and $\sigma(\feh)$.
Performing a Student's t-test
and the F-test on the $<\feh>$ and $\sigma(\feh)$ statistics which test
for the likelihood that two the samples have a common mean and variance
respectively, 
we find that they are inconsistent with one another at the
97$\%$ and 43$\%$ c.l.  We are now confident that the unweighted mean
$<\feh>$ (i.e.\ the metallicity of any given damped system) is decreasing with 
increasing redshift, but less certain that the scatter
in the [Fe/H] distribution decreases at high $z$. 

\begin{figure*}[ht]
\begin{center}
\includegraphics[height=5.8in, width=4.2in,angle=-90]{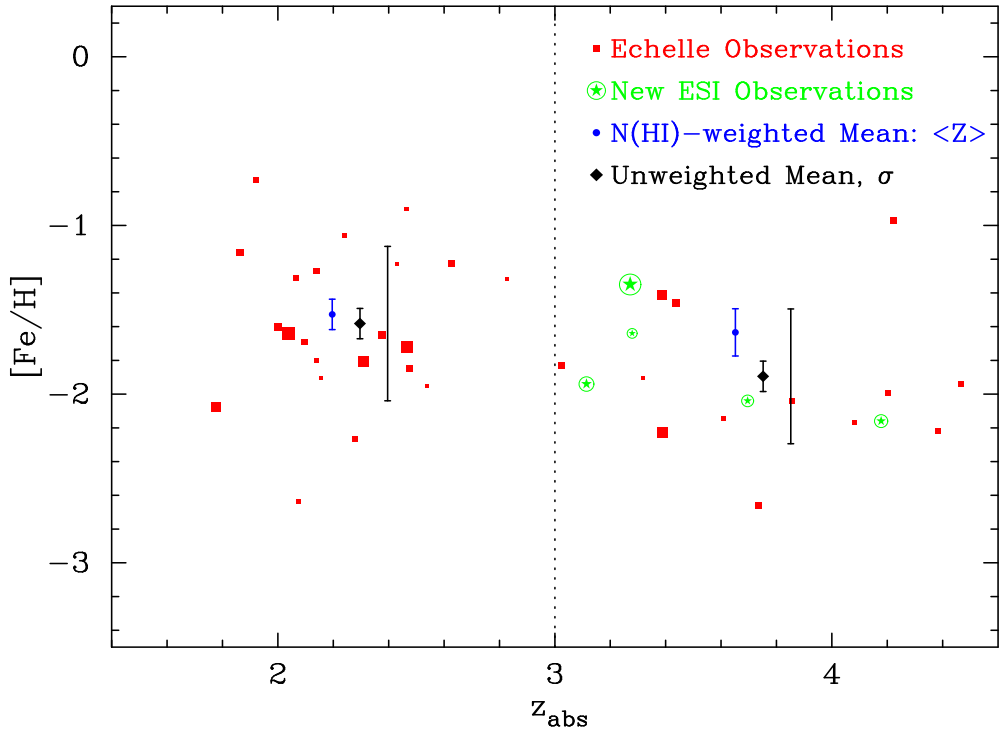}
\caption{[Fe/H] vs.\ $z_{abs}$ plot of the 40 damped \lya systems which 
comprise our entire $z > 1.7$ sample.  The squares indicate metallicity
measurements acquired from high resolution echelle observations and the stars
indicate the 5 new measurments obtained with the Echellette Spectrograph and
Imager.  For two samples ($z_{low} = [-1.8,3.0)$ and $z_{high} = [3.0, 4.7]$)
we present three statistics: (1) the HI-weighted mean metallicity $<Z>$ 
represented by the solid circle with bootstrap error;
(2) the unweighted mean logarithmic 
metallicity $<\feh>$ with bootstrap error depicted by the solid 
triangle; and (3) the scatter in the individual systems $\sigma(\feh)$
represented by the offset vertical error bar.
}
\end{center}
\label{fig:rslts}
\end{figure*}

The additional 7 systems
considered here nearly double the total HI content of the $z_{high}$ sample
and the new total ($H_T = 10^{22.1} \cm{-2}$) well exceeds the
$\N{HI}$ value measured for any single system.  Therefore, the results are
more robust to addition of individual systems.  For example, since the
publication of PW00 we have received criticism that the results were largely
biased by the single system towards PSS1443+22 ($z_{abs} = 4.23$;
$\feh \sim -1$).  Removing that data point from the current set of measurements
reduces the mean statistics ($<Z>_{high}$ by 0.1~dex and $<\feh>$ by 0.05)
but statistically the effect is only significant in the scatter as this 
data point dominates $\sigma(\feh)_{high}$.  Eliminating PSS1443+22 
reduces $\sigma(\feh)_{high}$ by 0.06 and a new F-test indicates the 
$z_{low}$ and $z_{high}$ distributions are inconsistent at $80\%$ c.l.
Perhaps more intriguing, however, is that
if one were to split the $z_{high}$ bin at $z=3.5$ into two sub-samples, 
eliminating the PSS1443+22 system would have the effect that $<Z>$ would
decrease markedly ($\sim 0.4$~dex) at $z>3.5$.  
We stress, however, that {\it there is no physical justification} for 
disregarding the system towards PSS1443+22; both the HI and Fe column 
densities are well measured and there may be a significant number of
high metallicity systems at $z>4$.  As our sample size increases significantly
in the upcoming year, we will further investigate smaller redshift intervals.

Along with the results from PW00, the new results imply an evolution in the
mean metallicity of neutral gas which is inconsistent with the majority of
chemical evolution models.  This includes the analyses of \cite{pei99},
\cite{edmns97}, and \cite{mny96} all of which predict a significantly 
larger decrease in $<Z>$ than observed.  \cite{prntz00} have noted that the
absence of evolution in $<Z>$ could be explained by a dust
obscuration scenario where no system with [Zn/H] + $\log \N{HI} > 21$
is observed.  While this model matches the observations well, these
authors did not incorporate their model within the formalism of \cite{fall93}
which is very likely to minimize the effects of dust obscuration.
There is an ongoing survey of a sample of radio-selected QSO's which will
help emperically assess the effects of dust obscuration \citep{ell00}.
Recently, \cite{mathlin00} presented a comprehensive chemical evolution
model for the damped \lya systems which included the \cite{press74}
cosmological formalism, the \cite{fall93} dust obscuration treatment,
and a detailed galactic chemical evolution scenario along the lines of
\cite{prntz00}.  Their results have the same failing as the majority of
previous models:  an underprediction of $<Z>$ and $<\feh>$ at $z>3$.
While they suggest that galaxy interactions and mergers may resolve this
discrepancy, it is unclear why these processes would not significantly alter
their results at $z < 3$.   \cite{cen99} have included
the effects of mergers and feedback in their numerical simulations
and their results indicate a milder evolution in $<Z>$ from $z=0-2$
for objects they identify as damped \lya systems. Unfortunately, this
treatment did not extend to $z=4$, but preliminary results appear
to agree with the results presented in this Letter (Cen \& Ostriker 2000;
priv.\ comm.).

In PW00, we noted that none of the systems with $\feh > -1.5$ have
large HI column density.  This could be explained by dust obscuration
\citep[e.g.][]{boisse98}, but it is also naturally explained by 
scenarios where a significant fraction of neutral gas has been converted
into stars as well as galactic models with a metallicity gradient and a
central HI hole \citep{wol98,efst00}.  The latter scenarios also   
account for the increasing scatter in [Fe/H] provided the gas begins to
be significantly consumed at $z<3$.
Examining Figure~3, one notes that none of the new systems
have $\feh < -2.5$ and the total sample $(N = 19$) has no system with 
$\feh < -2.7$~dex.  We emphasize that the ESI measurements are 
sensitive to metallicities well below $-3$~dex\footnote{Our 2$\sigma$
limits on the metallicity are
[Al/H]=~$-3.3$, [Fe/H]~$=-2.8$, and [O/H]~$=-4.0$ assuming the
feature spans 3 pixels with a SNR=20 and $\N{HI} = 20.6$}
and not a single system
has been removed from the analysis because no metal-lines were detected.  
While future measurements
may reveal a system or two with $\feh < -3$, we contend that there is a
physical lower limit to the metallicity of the damped \lya systems 
{\it at all epochs}.  Whether this lower limit on [Fe/H] is a function of
Pop~III pre-enrichment \citep[e.g.][]{wass00,ell00b}
or rapid metal enrichment is still a matter of 
debate.

We expect our efficiency on future nights with
ESI to be higher yielding at least 7 new [Fe/H] measurements
at $z>3$ per night.  Therefore, in the next two years our survey and
other similar ongoing projects (Sargent, Djorgovski; priv.\ comm.) will
reveal over 50 new measurements from ESI alone.  In fact, with one more
successful night we will have more systems in the $z_{high}$ sample than
the $z_{low}$ sample.  By pushing to lower magnitude limits $(R \sim 20$),
we will also be able to examine correlations between the metallicity
and quasar brightness. 

\acknowledgments

We acknowledge the very helpful Keck support staff for their efforts
in performing these observations. 
We would like to thank J. O'Meara and D. Tytler for help in acquiring data
on PH957. 
J.X.P. acknowledges support from a
Carnegie postdoctoral fellowship and thanks ESO for their hospitality
while this data was reduced. AMW was partially supported by 
NSF grant AST 0071257.

\end{document}